\documentclass[final,5p,times,twocolumn]{elsarticle}

\usepackage{amssymb}

\usepackage[utf8]{inputenc}
\usepackage{subfig}
\usepackage{siunitx}
\usepackage{amsmath}

\usepackage{multirow}
\usepackage{booktabs}
\usepackage{blindtext}
\usepackage{color}

\journal{Computational Materials Science}

\newcommand{\legline}[1]{\raisebox{-0.1cm}{\protect\includegraphics{line#1.pdf}}}

\begin{document}

\begin{frontmatter}



\title{Combined molecular dynamics and phase-field modelling of crack propagation in defective graphene}


\author[label1]{Arne Claus Hansen-Dörr}
\author[label2]{Lennart Wilkens}
\author[label2,label3]{Alexander Croy}
\author[label2,label3]{Arezoo Dianat}
\author[label2,label3]{Gianaurelio Cuniberti\corref{cor1}}
\author[label1,label3]{Markus Kästner\corref{cor2}}

\address[label1]{Institute of Solid Mechanics, TU Dresden, Dresden, Germany}
\address[label2]{Institute of Materials Science, TU Dresden, Dresden, Germany}
\address[label3]{Dresden Center for Computational Materials Science (DCMS), TU Dresden, Dresden, Germany}
\cortext[cor1]{gianaurelio.cuniberti@tu-dresden.de}
\cortext[cor2]{markus.kaestner@tu-dresden.de}

\begin{abstract}
In this work, a combined modelling approach for crack propagation in defective graphene is presented. Molecular dynamics (MD) simulations are used to obtain material parameters (Young's modulus and Poisson ratio) and to determine the energy contributions during the crack evolution. The elastic properties are then applied in phase-field continuum simulations which are based on the Griffith energy criterion for fracture.
In particular, the influence of point defects on elastic properties and the fracture toughness are investigated. For the latter, we obtain values consistent with recent experimental findings.
Further, we discuss alternative definitions of an effective fracture toughness, which accounts for the conditions of crack propagation and establishes a link between dynamic, discrete and continuous, quasi-static fracture processes on MD level and continuum level, respectively.
It is demonstrated that the combination of MD and phase-field simulations is a well-founded approach to identify defect-dependent material parameters.
\end{abstract}


\end{frontmatter}




\section{Introduction}
\label{sec:intro}
The multifunctionality of graphene is attractive for a large number of applications \cite{Zhu_Graphene_2010}.
It is an integral part of many composites \cite{Young2012,liu12} and serves as a prototypical example for other 2D materials \cite{Miro2014}.
Some of the beneficial properties are high thermal conductivity, high electron mobility at room temperature,
large surface area, high \textsc{Young}'s modulus, and good electrical conductivity.
However, one of the most important problems for the application of graphene in thermo/electrical
devices lies in the preparation of high-quality and well-defined specimen in bulk quantities.
Typically, the resulting atomic structure of graphene contains different structural deficiencies such as vacancy defects,
bond rotations, dislocation edges, grain boundaries, layer stacking and cracks \cite{Banhart2011}. Those defects can have a decisive impact on
the formation and propagation of cracks in graphene and related materials.
Computational methods have been very
helpful for bridging the scales from atoms to microstructures and for predicting structure-property
relationships \cite{Buehler2006}. For graphene, several atomistic studies -- mostly based on classical molecular dynamics (MD) simulations -- of crack propagation have been performed to extract elastic properties and fracture toughness \cite{omeltchenko97,Terdalkar2010,kim2012,Cao2013,Moura2013,le2014,budarapu15,Fan2017}. Only recently, the latter was measured
using \textit{in situ} tensile testing of suspended graphene \cite{Zhang2014}.

In general, understanding the propagation of cracks is important for improving the reliability of macroscopic structures and devices. Inherently, this issue is of multi-scale nature: cracks are formed by breaking chemical bonds, they propagate in the host material and ultimately reach macro-scale dimensions. Additionally, linking atomistic and continuum scales is very challenging due to the dynamic nature of crack propagation \cite{Buehler2006}. During this process, the work of applied external loads is converted into surface energy (crack), potential energy (deformations and defects) and kinetic energy (movement of atoms). In a real material the latter is typically dissipated as heat. On the other hand, energetic criteria motivated from a continuum perspective, such as the one by Griffith \cite{griffith_phenomena_1921}, assume that the work is solely converted into
surface energy after elastic deformation happened. The critical energy release rate, which is needed to increase the surface of the crack, is the basis for continuum
descriptions of fracture \cite{kuhn_continuum_2010,miehe_phase_2010}. However, a fully coupled continuum model accounting for various energy conversion effects is computationally quite expensive and still needs verification and validation for all material parameters. This is the motivation to introduce an effective fracture toughness, which depends on the conditions of crack propagation altogether and thus drastically reduces the number of required material parameters.

Computational continuum methods, such as the finite element method (FEM) combined with a phase-field (PF) model,
which are based on energetic criteria have successfully been used to model crack initiation and propagation in different materials. In the first approaches,
brittle fracture has been studied \cite{bourdin_variational_2008,miehe_thermodynamically_2010} and was later extended to ductile fracture for small \cite{ambati_phase-field_2015}
and finite plastic strains \cite{miehe_phase_2016-2,ambati_phase-field_2016-1}. The transition
from brittle to ductile failure has been analysed \cite{miehe_phase_2015-1}, too. Recent developments include crack propagation in heterogeneous materials \cite{hossain_effective_2014,schneider_phase-field_2016,nguyen_phase-field_2016}.
To obtain the material parameters entering the continuum approaches, the combination
with molecular dynamics simulations has been successfully applied.
For example, crack propagation has been studied in b.c.c.\ crystals using MD/FEM \cite{kohlhoff_crack_1991}
and in aragonite using MD/PF \cite{patil_comparative_2016, padilla_3d_2017}.

In this article, elucidating the relation between energy conversion and effective fracture toughness,
we use a combination of a continuum PF model with MD simulations
to describe crack propagation in defective graphene sheets. The MD results are used
to obtain the materials properties -- \textsc{Young}'s modulus, \textsc{Poisson} ratio,
regularization length and fracture toughness -- required for the PF method in dependence on
the defect density. The material parameters for the continuum simulations are determined by MD simulations using homogenization and crack propagation approaches. While the MD scale resolves every single atom and its bonds, the PF simulations are based on a homogeneous continuum. In the present contribution, only mode-I cracks are considered without loss of generality. We discuss in detail how the effective fracture toughness can be extracted from the simulations.

The article is structured as follows: in Sec.\ \ref{sec:theory} we give
an overview of the two methods, MD and PF, used to study the crack propagation in graphene sheets. The results of the respective simulations are presented and discussed in Sec.\ \ref{sec:paramident}. Finally, we summarize our results and discuss their implications on crack propagation in Sec.~\ref{sec:conclusion}.


\section{Methods}
\label{sec:theory}
\subsection{Molecular dynamics}
On an atomistic level the initiation and propagation of cracks in a material involves the breaking
of chemical bonds. While ab initio methods, such as density functional theory, are able to
account for bond breaking and formation, they are computationally expensive for very large systems \cite{MMMN2016}.
On the other hand, classical molecular dynamics simulations are able to describe setups with millions of atoms.
For treating crack propagation and to enable an atom to change its coordination within a purely classical
simulation, so-called bond-order potentials can be employed \cite{abell85,tersoff88,Brenner90,Brenner2002,pastewka_mrovec_moseler_gumbsch_2012}. Such potentials can be written in the form
\begin{equation}
E = \frac{1}{2}\sum_{i,j} \left[ V_{\text{rep}}(R_{ij}) + b_{ij} V_{\text{attr}}(R_{ij})\right]\;,
\end{equation}
where $R_{ij}$ is the distance between two atoms, $V_{\text{rep}}$ and $V_{\text{attr}}$ denote repulsive and attractive potentials,
respectively, and $b_{ij}$ is the bond-order parameter which contains the dependence on the coordination of the atoms $i$ and $j$.
The strength of the bond thus depends on the chemical environment which allows it to mimic chemical reactions to a certain
extend and to describe fracture in solids \cite{Tersoff89,pastewka_mrovec_moseler_gumbsch_2012,Harrison2015}.

In order to atomistically describe the crack propagation in graphene sheets, we use classical molecular dynamics
simulations implemented in LAMMPS \cite{plimpton95}. All fracture simulations are performed employing the
long-range bond-order potential for graphene (LCBOP) of Ref.~\cite{los03}. The setup and the initialization of the simulations
are described in Sec.~\ref{sec:methods_link}. Since we are also interested in the energy contributions arising from the crack, we
use atom-number, volume and energy conserving (NVE) ensembles.
Additionally, we perform calculations with a short-ranged Tersoff potential~\cite{tersoff94}
which has been parametrized for carbon as a comparison.
\subsection{Continuum phase-field model}
\begin{figure*}[t]
	\centering
	\subfloat[Discrete crack]{
		\includegraphics{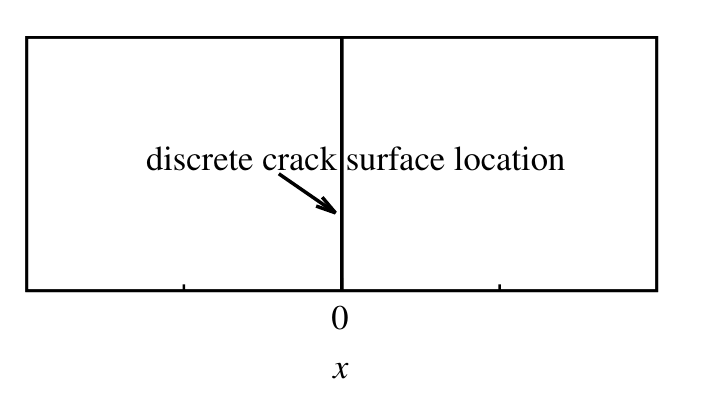}
	}
	\subfloat[Regularized phase-field crack]{
		\includegraphics{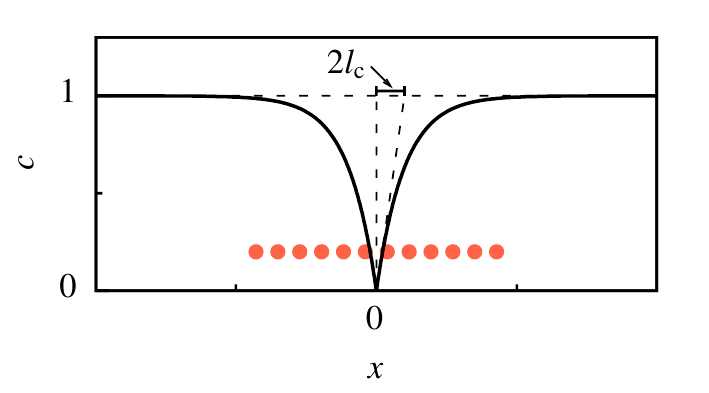}
	}
	\caption{Different crack representations are compared for the one-dimensional case. On the left, a discrete crack surface is located at $x=0$ within an infinitely long domain, i.e. $x\in(-\infty,\infty)$. On the right, the discrete crack is regularized and now represented by a phase-field $c$ smoothly bridging the fully intact ($c=1$) and fully broken ($c=0$) state. The parameter $l_\text{c}$ controls the width of the transition zone. The salmon spheres foreshadow the distance between the atoms compared to the phase-field length scale, which is of the same order.}
	\label{fig:discSmeared}
\end{figure*}
\begin{figure}[t]
	\centering
	\includegraphics{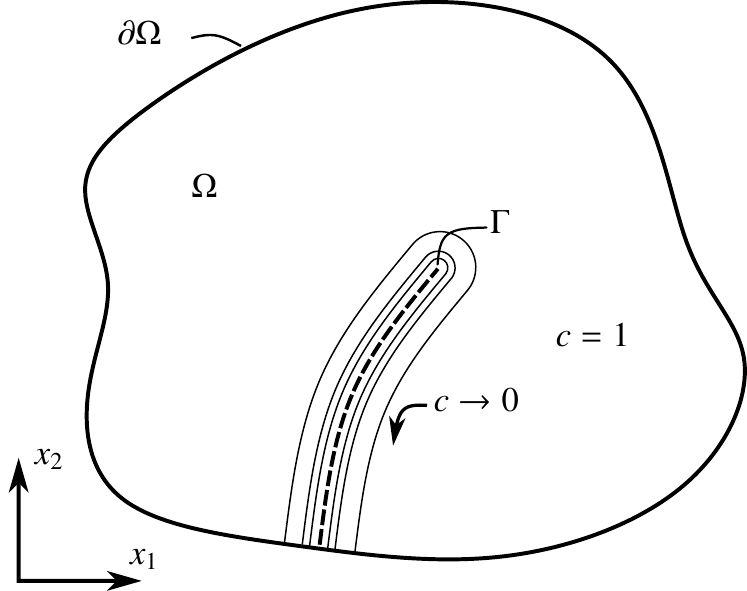}
	\caption{Schematic illustration of a formerly discrete crack surface $\Gamma$ (dashed line), which is regularised using a phase-field $c$. Far away from the crack, $c=1$. The contour lines depict the steep gradient of the phase-field. }
	\label{fig:2DPhaseField}
\end{figure}
On the continuum level, the description of crack propagation is based on the energetic cracking criterion which goes back to the work of Griffith~\cite{griffith_phenomena_1921}. Accordingly, the energy release rate $\mathcal{G}$ can be written as
\begin{equation}
\label{eq:GriffithCrit}
\mathcal{G}=-\frac{\text{d}\Pi}{\text{d}\Gamma}\,,
\end{equation}
where $\Pi$ is the energy of the system and $\Gamma$ is the crack surface. In other words, the rate of energy released depends on how much of the system energy is needed for a hypothetical increase of the crack surface. The crack propagates if $\mathcal{G}\geq\mathcal{G}_\text{c}$ where $\mathcal{G}_\text{c}$ is a material parameter and is referred to as the critical energy release rate. For certain cases like pure mode-I cracks as considered within this work, the value $\mathcal{G}_\text{c}$ can directly be related to the fracture toughness $K_\text{Ic}$~\cite{gross_bruchmechanik_2011}.\par
The energy for an elastic solid incorporating cracks reads
\begin{equation}
\label{eq:freeHelmdisc}
\Pi=\Psi^\text{el}+\mathcal{G}_c\int  \,\text{d}\Gamma\,.
\end{equation}
The second term on the right hand-side captures the surface energy of the crack. This functional is the basis for a numerical implementation. Typical finite element discretizations use a conforming mesh to represent the crack path $\Gamma$, i.e.\ the analysis mesh has to be updated during crack propagation. Different from such a discrete crack representation, phase-field models solve an additional scalar field problem for the phase-field order parameter $c\in[0,1]$ representing a regularised crack topology illustrated for the one-dimensional case with a crack at $x=0$ in Fig.~\ref{fig:discSmeared}. The so-called phase-field is coupled to the mechanical boundary value problem and allows for the proper modelling of crack initiation and propagation. In this approach, topological updates of the analysis mesh are avoided. \par
\begin{equation}
\label{eq:freeHelm}
\begin{split}
\Pi\left[\boldsymbol{\varepsilon},c\right]&=\underbrace{\int_{\Omega}(c^2+\eta)\,\psi^\text{el}(\boldsymbol{\varepsilon})\,\text{d}\Omega}_{\Psi^\text{el}}\\\
&+\underbrace{\mathcal{G}_c\int_{\Omega} \frac{1}{4l_\text{c}}\left[(1-c)^2 + 4 l_\text{c}^2 \nabla c \cdot \nabla c \right]\,\text{d}\Omega}_{\Pi^\text{c}}~\text{.}
\end{split}
\end{equation}
The crack energy $\Pi^\text{c}$ implements the energetic criterion mentioned above in a regularised manner, i.e. the integral over a sharp crack in Eq.~\eqref{eq:freeHelmdisc} is replaced by its regularised approximation where the kernel of the integral can be interpreted as a crack surface density $\gamma(c)$~\cite{miehe_thermodynamically_2010}.
Fig.~\ref{fig:2DPhaseField} illustrates the phase-field of a formerly discrete crack surface $\Gamma$ for 2D. \par
In this work, only small strains $\boldsymbol{\varepsilon}$ are considered.
Hence, the elastic energy per unit volume can be written as $\psi^\text{el}=\tfrac{1}{2}\boldsymbol{\varepsilon}:\boldsymbol{E}:\boldsymbol{\varepsilon}$, with the fourth order elasticity tensor $\boldsymbol{E}$. In regions of cracked material, the elastic energy density is degraded by multiplying $\psi^\text{el}$ with the degradation function $c^2$, cf. Eq.~\eqref{eq:freeHelm}, i.e. the material looses its integrity in regions where the regularised crack develops. A discussion of different degradation functions can be found in~\cite{kuhn_degradation_2015}. For reasons of numerical stability, a certain residual stiffness $\eta=10^{-7}$ is maintained.\par
In the following, the specimen is loaded using monotonically increasing displacement boundary conditions and the crack will propagate under mode-I. This implies that no special treatment for the irreversibility of crack growth \cite{linse_convergence_2017} is needed and cracking under pressure will never occur. The latter justifies the choice of fully degrading the elastic energy $\psi^\text{el}$, i.e.\ no particular split as in~\cite{miehe_phase_2010,amor_regularized_2009,steinke_phase-field_2018} is incorporated.\par
The Euler-Lagrange equations of the coupled boundary value problem can be derived in a variational manner~\cite{miehe_thermodynamically_2010,miehe_phase_2010}. The resulting weak form is discretized using locally refined Truncated Hierarchical B-splines (THB-splines)~\cite{hennig_bezier_2016,hennig_bezier_2017,hennig_thb_splines_2018}, which allows for efficient computations because the steep gradients arising from the phase-field can be resolved locally using adaptive mesh refinement. The non-linear coupled equations are solved using a monolithic scheme with a heuristic adaptive time-step control.\par 
It is important to note, that the value of $\mathcal{G}_\text{c}$ which is applied in the PF simulations experiences a numerical influence
\begin{equation}
\label{eq:gcnum}
\mathcal{G}_\text{c}^\text{num}=\mathcal{G}_\text{c}\left(1+\frac{h}{4l_\text{c}}\right)\,,
\end{equation}
where $h$ is a characteristic length of the finite element mesh near the crack~\cite{bourdin_variational_2008}. This yields a slight overestimation of the crack energy, i.e.\ the numeric value $\mathcal{G}_\text{c}^\text{num}$ is slightly higher than the specified $\mathcal{G}_\text{c}$. If not stated differently, the characteristic element size is $h = \SI{0.25}{\angstrom}$ and the numerical influence is compensated for all phase-field simulations.

\subsection{Linking both methods}\label{sec:methods_link}
In the present study, four parameters, the \textsc{Young}'s modulus~$E$, the \textsc{Poisson} ratio~$\nu$, the regularization length~$l_\text{c}$ and the critical energy release rate $\mathcal{G}_\text{c}$ are required for a complete continuum description. Those material parameters
are obtained from MD simulations and then applied to continuum PF simulations and the results are quantitatively compared.

\begin{figure*}[p]
	\centering
	\subfloat[Bidirectional tension test setup]{\label{fig:elastHomo_a}
		\includegraphics[trim={0.0cm -0.8cm 0.7cm 0cm},clip]{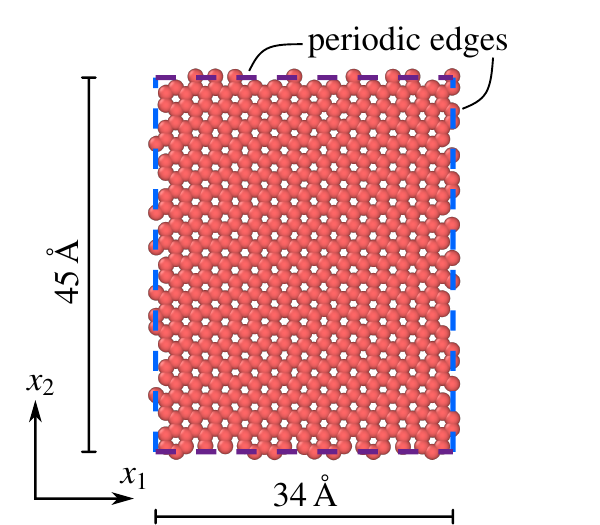}
	}
	\subfloat[\textsc{Young}'s moduli vs defects]{
		\includegraphics{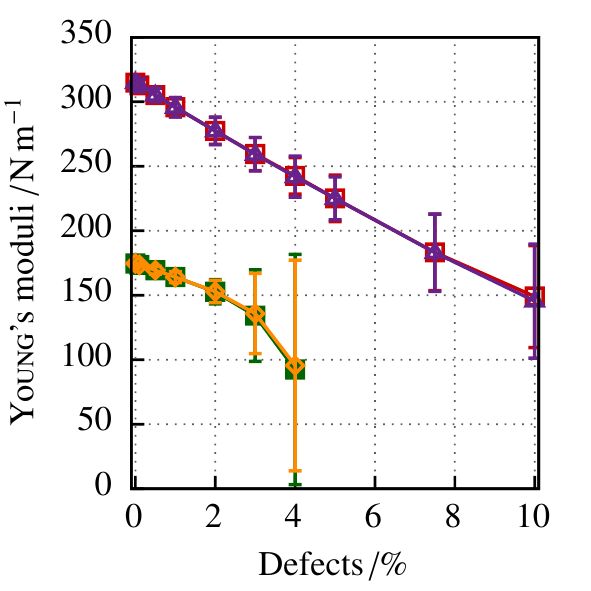}
	}
	\subfloat[\textsc{Poisson} ratios vs defects]{
		\includegraphics{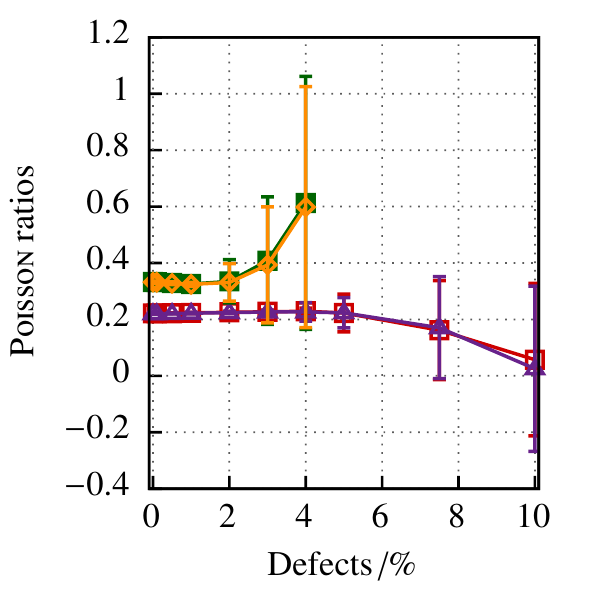}
	}\\
	\begin{tabular}{c l|c l}
		\multicolumn{2}{c|}{Load in $x_1$ direction} &\multicolumn{2}{c}{Load in $x_2$ direction}\\
		\legline{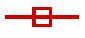}& LCBOP &\legline{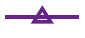} &LCBOP \\
		\legline{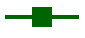}& Tersoff-1994 &\legline{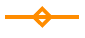}& Tersoff-1994 \\
	\end{tabular}
	\caption{The material parameters and their specific uncertainty of single layer graphene sheets are highly dependent on the defects. The bidirectional tension test setup for the homogenization of a defect-free sheet is depicted on the left hand-side (a).  The diagrams (b)--(c) show the dependence of the material parameters on the defects. Error bars indicate the 68.3\% quantile. For the continuum simulations, isotropic material is assumed which can be justified by the equality of both loading directions.}
	\label{fig:elastHomo}
	\vspace{0.8cm}
	\hrule
\end{figure*}

The two elastic material parameters, $E$ and $\nu$, are determined in a computational homogenization scheme. Fig.~\ref{fig:elastHomo_a} describes the general simulation setup. Graphene sheets of size $34 \times \SI{45}{\angstrom}^2$ with periodic boundary conditions are first fully relaxed in the MD simulation.
Then, the systems are stretched and compressed in one direction by $0.1\%$, respectively. From the energy differences, the \textsc{Young}'s modulus is estimated in that direction. The \textsc{Poisson} ratio is obtained
from the resulting lateral change of the simulation box. In the presence of defects, $500$ different defective configurations for each defect concentration have been used and the resulting elastic constants are averaged over these realizations. There are different defect types in graphene such as Stone-Wales defects, single/multiple vacancies and dislocation-like defects~\cite{Banhart2011}.
For simplicity, this work focuses on point defects within graphene sheets, i.e.\ multiple atoms are randomly removed from a perfect sheet.

\begin{figure*}[p]
	\centering
	\subfloat[One sample of $200$ MD setups]{
		\includegraphics[trim={0cm 0cm 9.27cm 0cm},clip]{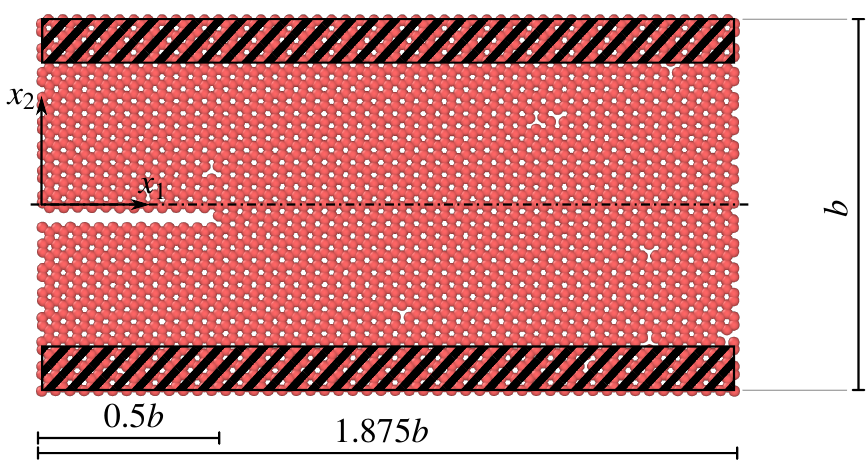}
	}\hspace*{-0.292cm}
	\subfloat[Continuum setup using a PF model]{
		\includegraphics[trim={8.79cm 0cm 0cm 0cm},clip]{fig4ab.pdf}
	}
	\caption{The crack propagation simulations are carried out using two approaches featuring the same domain sizes and -- as far as applicable -- equal boundary conditions. On the left (a), one of $200$ MD setups is depicted with 2\% defects. A so-called surfing boundary condition, cf.\ Eq.~\eqref{eq:surfBCFormulax}~\&~\eqref{eq:surfBCFormulay}, is applied to the atoms in the black hatched areas. The crack is expected to form a wavy path. On the right (b), the continuum setup using a PF model is depicted. The initial crack is incorporated by setting the PF to zero. In contrast to the MD simulations, the surfing boundary condition is not applied to a surface but the upper and lower black marked edge of the specimen. It is noted, that for the MD setups the crack cannot be exactly at $x_2\equiv0$ which is due to the atomistic, discontinuous structure.}
	\label{fig:simSetups}
	\vspace{0.8cm}
	\hrule
\end{figure*}
\begin{table*}[p]
	\centering
	\caption{Summary of the elastic parameter averages for the \textsc{LCBOP} potential depicted in Fig.~\ref{fig:elastHomo} as applied in the continuum simulations. The average values for the $x_1$- and $x_2$-direction are averaged again to get a single value for each parameter, i.e. it is reasonable to assume isotropic elasticity. The fracture toughness shown in the last two rows is calculated using the approaches presented in Sec.~\ref{sec:1stAlternative} and Sec.~\ref{sec:2ndAlternative}.}
	\label{tab:elastParam}
	\begin{tabular}{l l c c c c c c c}
		\toprule
		Defects\,/\% & & 0 & 0.1 &0.5 &1 &2 &3 & 4\\ \midrule
		$E/\si{\newton\per\meter}$ & & 314.88 & 312.70&305.21 &295.77 &277.44 &259.39 & 242.25\\
		$\nu$ & &0.22090 &0.22095 &0.22167 &0.22261 &0.22471 &0.22678 &0.22794 \\ \midrule
		$\mathcal{G}_\text{c}/\si{\nano\newton}$ & using $\Delta E_\text{min}$ & 5.378&5.348 &5.054 &4.586 & 3.546&2.416 &1.454 \\
		$\mathcal{G}_\text{c}/\si{\nano\newton}$ & using $\Delta E_\text{tot}$ & 23.896&23.612 &23.499 &23.385 & 22.658&22.669 &22.296 \\ \bottomrule
	\end{tabular}
\end{table*}

\subsubsection*{Determination of fracture material parameters}
Fig.~\ref{fig:simSetups} describes the computational domains for the MD and continuum crack simulations. The width of the specimen is set to $b=\SI{6.4}{\nano\meter}$. Both domains have an initial crack, which is given in terms of removed atoms in the MD case. Due to the atomistic structure, the initial crack is slightly asymmetrical, which did not have an influence on the obtained crack patterns. For the continuum simulation, the PF is set to $c=0$ along the initial crack.\par
As already mentioned above, the fracture toughness can be related to the critical energy release rate in linear fracture mechanics, cf.~\cite{gross_bruchmechanik_2011}: Exploiting the linear relation $\mathcal{G}_\text{c}=(1-\nu)^2 K_\text{Ic}^2/E$ for plane strain and mode-I cracks, as considered within this work, the determination of the fracture toughness is equivalent to determining $\mathcal{G}_\text{c}$. Due to this equivalence, the general, conceptual difference between both quantities is disregarded from now on and $\mathcal{G}_\text{c}$ is referred to as fracture toughness, too, for the sake of brevity and readability.\par 
The task of finding a defect-dependent value for the fracture toughness within the MD simulations, which is provided to the continuum scale, is not straight-forward as can be seen from the following two challenges:
\begin{enumerate}
	\item A stable crack propagation in MD and continuum simulations is necessary to be able to observe various effects, and
	\item the quantitative estimation of the fracture toughness from the MD simulations remains to be discussed, because the quasi-static continuum phase-field model has to account for discrete, dynamic fracture processes.
\end{enumerate}
The first issue is highly dependent on the choice of the boundary condition: In this work, a so-called surfing boundary condition~\cite{hossain_effective_2014} 
\begin{align}\label{eq:surfBCFormulax}
u_1&\equiv 0\,,\\\label{eq:surfBCFormulay}
u_2(x_1,x_2,t)&=\text{sgn}(x_2)\frac{A}{2}\left[ 1-\tanh\left( \frac{x_\text{IP}}{d}   \right)  \right]\,,\\
\text{where}&\quad x_\text{IP}=x_1-x_0-v\cdot t\,,\notag
\end{align}
is applied to the upper and lower edges of the MD and continuum domain, cf.\ Fig.~\ref{fig:simSetups}, that smoothly propagates the crack in time and allows for detailed observations and good comparability between the MD and continuum simulations. While the horizontal displacement remains fixed, see Eq.~\eqref{eq:surfBCFormulax}, the vertical displacement follows a hyperbolic tangent, where the inclination point position $x_\text{IP}$ travels in positive $x_1$-direction, Eq.~\eqref{eq:surfBCFormulay}. The shape-governing parameters are chosen $A=0.13b$ and $d=0.5b$ for both MD and continuum simulations. The initial inclination point position $x_0$ lies between $-1.6b\dots-b$ to ensure a smooth increase of loading. The exact position within the given interval did not have an influence. For the MD simulations, the velocity is set to $v=\SI{2.56}{\nano\meter\per\pico\second}$, which yields quasi-static surfing boundary conditions compatible with the phase-field description and thus justifies the comparison to quasi-static continuum simulations, in which the velocity is irrelevant. The choice of this type of boundary condition enables stable crack propagation and the use of a monolithic solution strategy for the continuum simulations. Additionally, it smoothly propagates the crack along the $x_1$-axis and justifies the use of a quasi-static modelling approach. \par
\begin{figure*}[t]
	\centering
	\subfloat[Resulting crack path with arc length and effective length of the crack path pointed out]{	\label{fig:ensembleAverageDemo_a}
		\includegraphics[trim={0cm -1.3cm -0.4cm 0cm},clip]{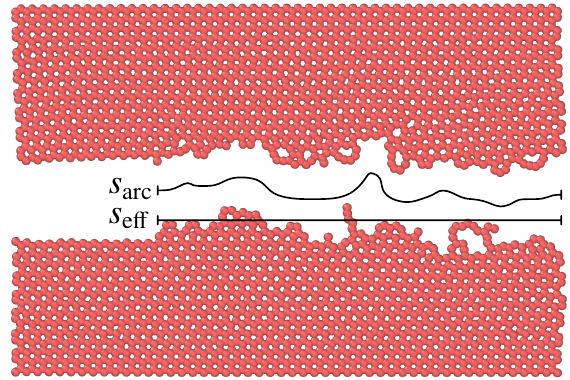}
	}
	\subfloat[Vertical reaction force]{
		\includegraphics{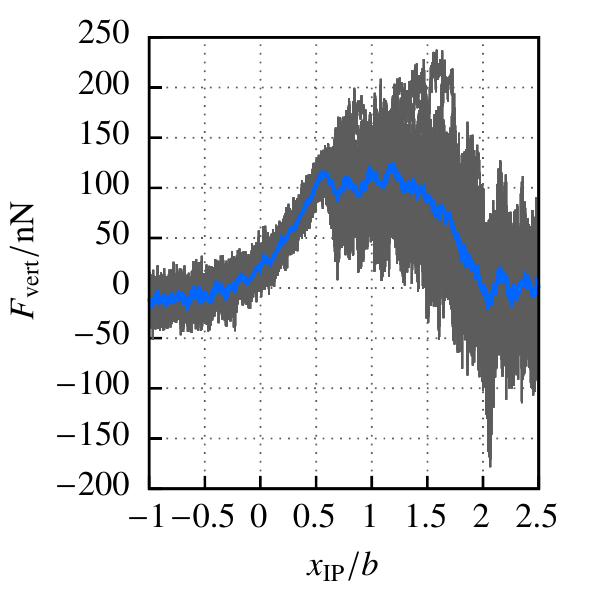}
	}
	\subfloat[Total energy]{
		\includegraphics{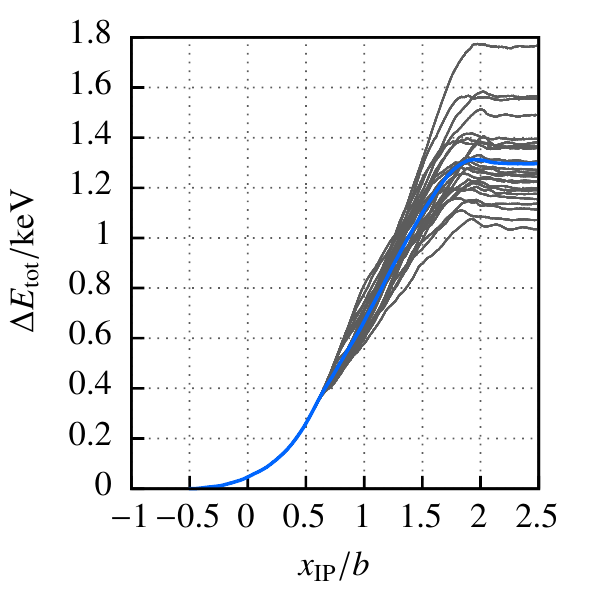}
	}
	\caption{On the left (a), one of $250$ resulting crack paths is depicted. Regarding the length increase $\Delta\Gamma$ of the crack, two values, the arc length $s_\text{arc}$ and the effective length $s_\text{eff}$, can be defined, which is of crucial importance for the determination of the fracture toughness. The two diagrams show the vertical reaction force (b) and total energy (c) of a perfect single layer sheet. All $250$ samples with slightly different initial conditions are depicted as grey lines in the background. The blue lines resemble the ensemble average for every $x_\text{IP}(t)$. $F_\text{vert}$ is the average of the absolute values of the two forces acting on the two black striped areas in Fig.~\ref{fig:simSetups} on the left hand-side. The initial negative offset of $F_\text{vert}$ is due to the initial random perturbation of the atoms which results in small reaction forces. }
	\label{fig:ensembleAverageDemo}
\end{figure*}

The second issue concerns the second term on the right hand-side of Eq.~\eqref{eq:freeHelmdisc}: The crack energy $\Pi^\text{c}$ is interpreted as a dissipated energy $\mathcal{D}$. Thus, during an increase of the crack surface $\Delta\Gamma$, $\Delta\mathcal{D}=\mathcal{G}_\text{c}\Delta\Gamma$ yields the increase of the dissipated energy. An appropriate measure for a finite increase of the dissipated energy $\Delta\mathcal{D}$ and the dedicated crack surface increase $\Delta \Gamma$ would enable the determination of the unknown fracture toughness
\begin{equation}
\label{eq:finiteGriffithCrit}
\mathcal{G}_\text{c}=\frac{\Delta\mathcal{D}}{\Delta\Gamma}\,.
\end{equation}
Since an incremental evaluation of Eq.~\ref{eq:finiteGriffithCrit} is quite difficult for the MD simulations, two clearly defined states, the beginning and the end (fully broken state) of a simulation are picked to evaluate the equation.
As mentioned above, the MD simulations produce wavy crack patterns, while the PF does not. This raises the question how the crack surface, crack length in two dimensions, is determined. Hossain~\cite{hossain_effective_2014} has shown, that a higher fracture toughness is observed for a meandering crack path $s_\text{arc}$, which travels in $x_1$-direction, cf. Fig.~\ref{fig:ensembleAverageDemo_a}, compared to a straight crack of length $s_\text{eff}$ in $x_1$-direction. This underlines the importance of the choice of $s_\text{arc}$ or $s_\text{eff}$.
For the present case, the effective distance in $x_1$-direction between the right end and the crack tip $s_\text{eff}=1.375b=\SI{8.8}{\nano\meter}$ ($= \text{right end of the specimen}-\text{initial crack tip position}$) has to be used as finite increase $\Delta\Gamma$ and not the (longer) arc length $s_\text{arc}$ of the wavy crack path, because the continuum approach, where the fracture toughness is applied, will give straight crack paths, since the atomistic structure is not resolved. \par
For the finite increase of the dissipated energy $\mathcal{D}$ two different approaches, \textit{step-by-step energy minimization} and \textit{total energy difference}, are considered. \par
\begin{figure*}[p]
	\centering\subfloat[$x_\text{IP}=\SI{0}{\nano\meter}$]{
		\includegraphics[trim={4cm 5.8cm 3.0cm 5.5cm},clip,scale=0.20]{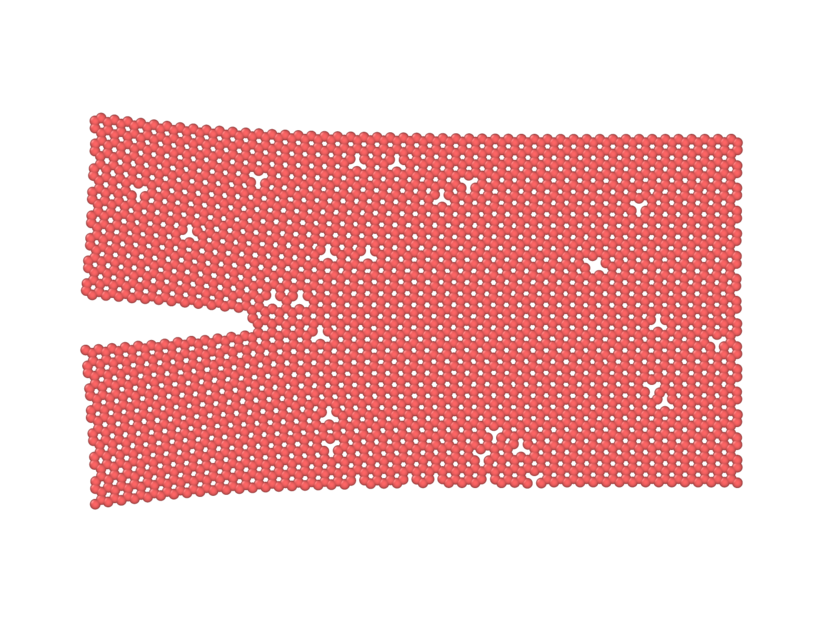}
		\includegraphics[trim={3.5cm 5.8cm 3.0cm 5.5cm},clip,scale=0.20]{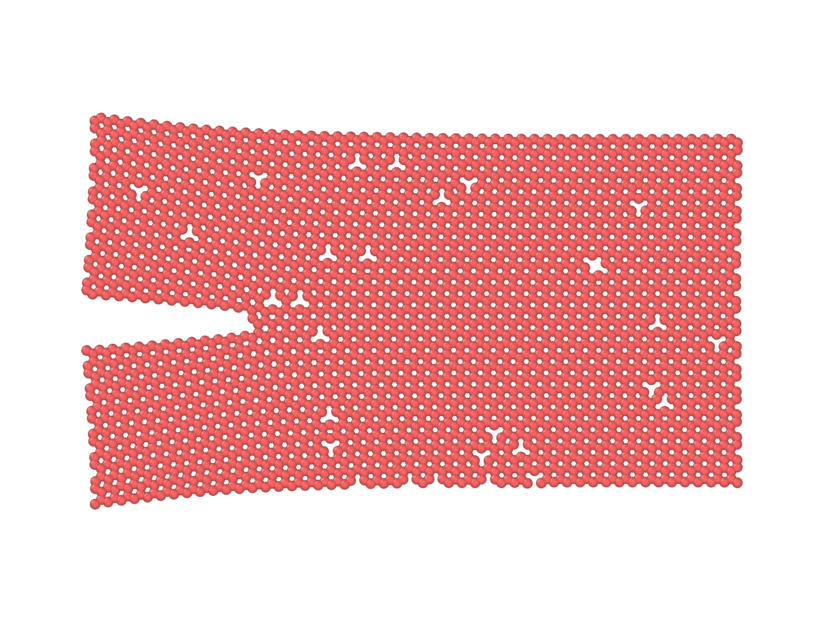}}\\
	
	\centering\subfloat[$x_\text{IP}=\SI{8.96}{\nano\meter}$]{
		\includegraphics[trim={4cm 4.5cm 3.0cm 4.5cm},clip,scale=0.20]{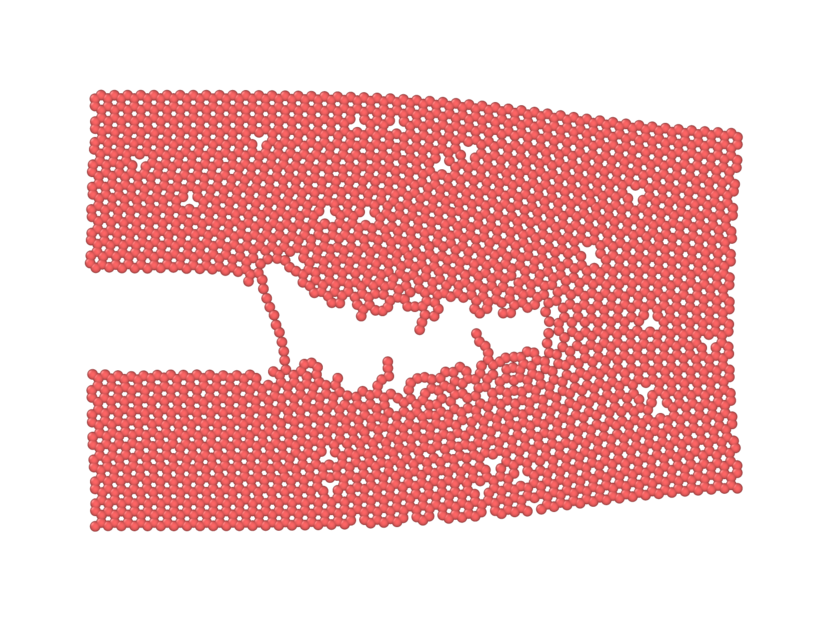}
		\includegraphics[trim={3.5cm 4.5cm 3.0cm 4.5cm},clip,scale=0.20]{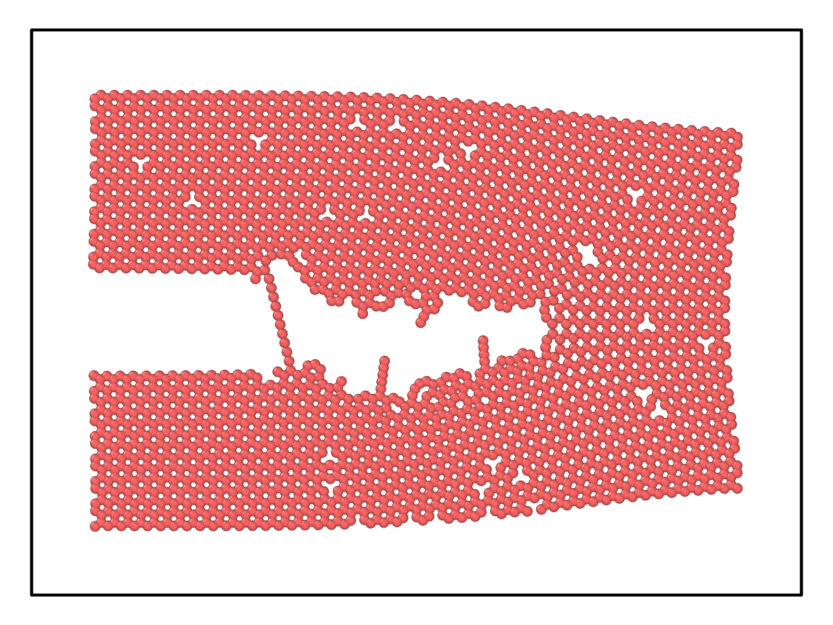}}\\
	
	\centering\subfloat[$x_\text{IP}=\SI{11.26}{\nano\meter}$]{
		\includegraphics[trim={4cm 4.5cm 3.0cm 4.5cm},clip,scale=0.20]{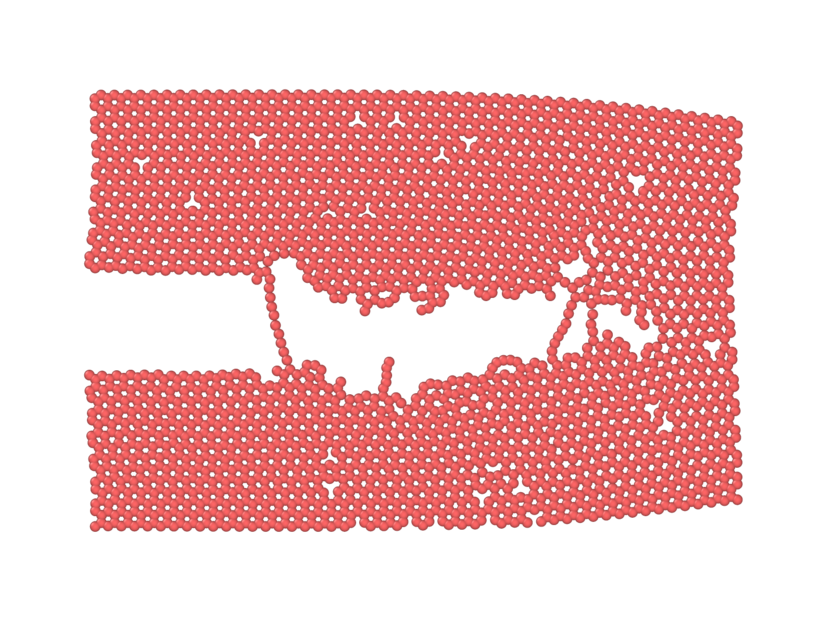}
		\includegraphics[trim={3.5cm 4.5cm 3.0cm 4.5cm},clip,scale=0.20]{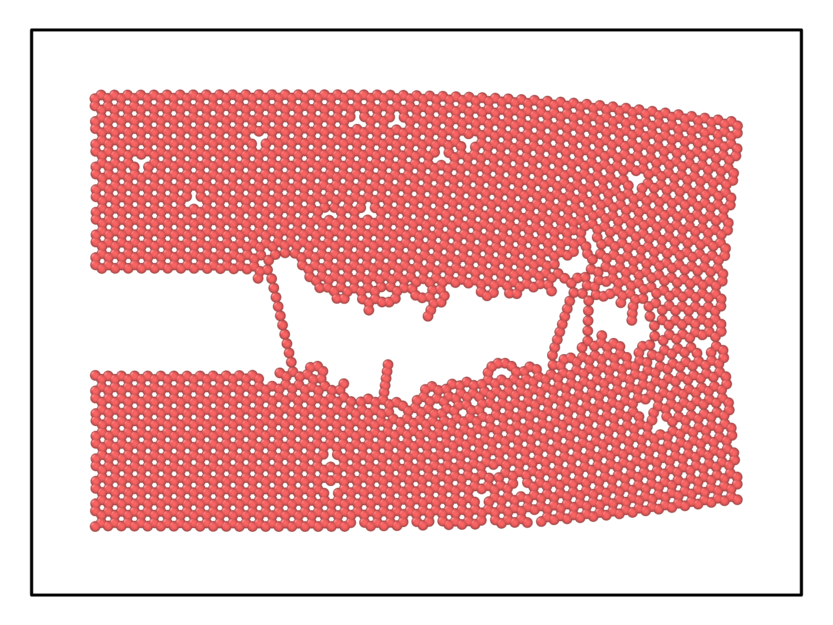}}\\
	
	\centering\subfloat[$x_\text{IP}=\SI{15.36}{\nano\meter}$]{
		\includegraphics[trim={4cm 4.5cm 3.0cm 4.5cm},clip,scale=0.20]{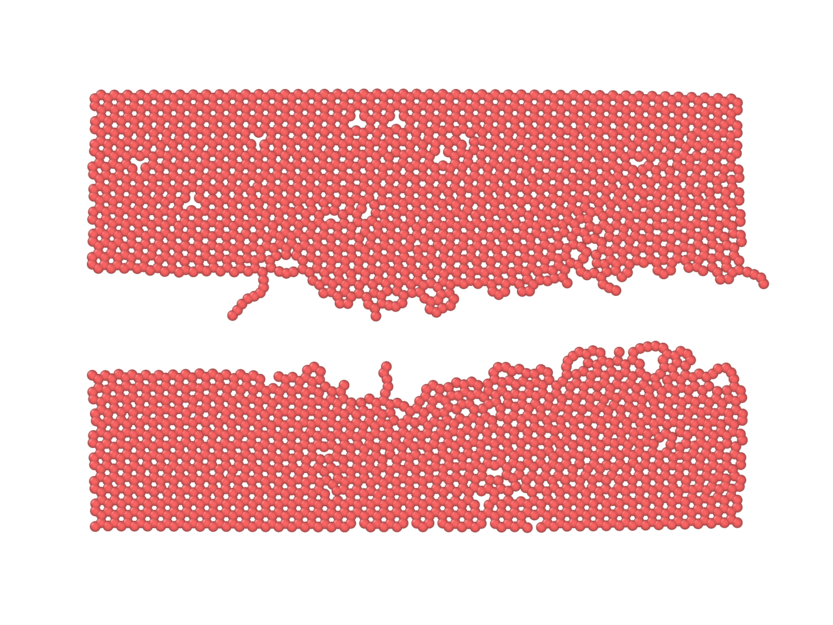}
		\includegraphics[trim={3.5cm 4.5cm 3.0cm 4.5cm},clip,scale=0.20]{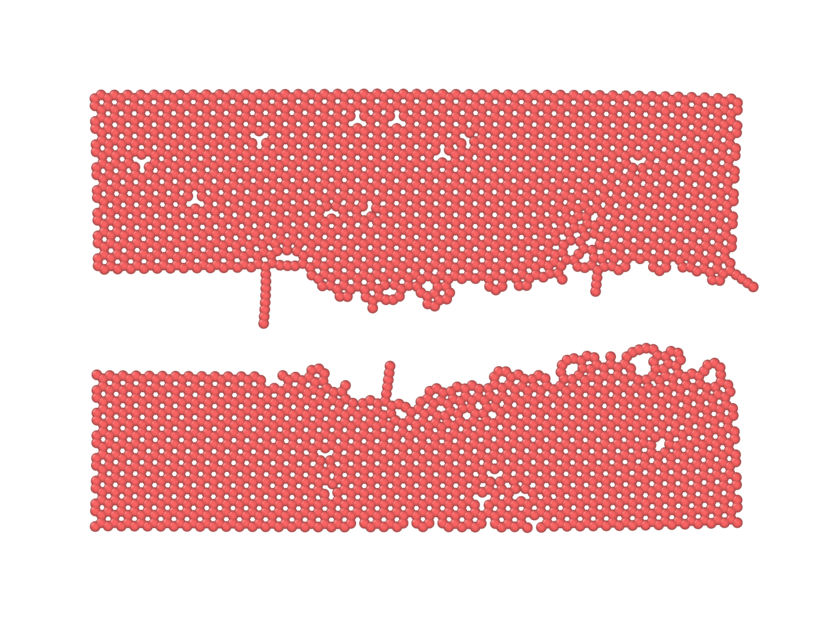}}\\
	
	\caption{Illustration of the cracked graphene single sheet for different positions (a)-(d) of the inclination point $x_\text{IP}$ before (on the left) and after (on the right) the minimization. The defect percentage is 1\%. Only small differences occur in the arrangement of the atoms, which is due to the minimization. }
	\label{fig:crackPathComp}
\end{figure*}

The first method estimates the dissipated energy from the broken bonds within the MD simulations: At certain points the simulation is paused and a minimization of the energy is performed. Compared to the initial minimized energy value $E_\text{min}^\text{start}$, the increase can be identified with the dissipated energy by breaking atomic bonds. Fig.~\ref{fig:crackPathComp} shows a comparison between the atomic configuration before and after the minimization for different inclination point positions $x_\text{IP}$. Minimization is also performed for the fully cracked specimen, which yields a value $E_\text{min}^\text{end}$. The fracture toughness is now estimated by
\begin{equation}
\label{eq:1stAltEst}
\mathcal{G}_\text{c}=\frac{\Delta E_\text{min}}{\Delta \Gamma} =\frac{E_\text{min}^\text{end}-E_\text{min}^\text{start}}{s_\text{eff}}\,,
\end{equation}
where $s_\text{eff}=\SI{8.8}{\nano\meter}$ is the effective crack length of the fully cracked material as discussed above. \par
The second method uses the value of the difference in total energy after complete rupture $\Delta E_\text{tot}$ at $x_\text{IP}=2$. The fracture toughness is estimated by
\begin{equation}
\label{eq:2ndAltEst}
\mathcal{G}_\text{c}=\frac{\Delta E_\text{tot}}{\Delta \Gamma} =\frac{E_\text{tot}^\text{end}-E_\text{tot}^\text{start}}{s_\text{eff}}\,.
\end{equation}
This implies, that the complete irreversible energy (temperature increase, broken bonds) in the MD approach contributes to the fracture toughness in the continuum simulations. Below in Sec.~\ref{sec:paramident}, the performance and validity of both approaches are discussed.\par
In contrast to Ref.\ \cite{patil_comparative_2016}, the regularization length is set to a value in the order of the graphene-graphene bond-length, cf.\ Fig.~\ref{fig:discSmeared} on the right, and takes the value $l_\text{c}=\SI{0.96}{\angstrom}$. The validity of this approach is discussed later in Sec.~\ref{sec:lengthScaleDisc}.


\begin{figure}[t]
	
	\centering
	\includegraphics{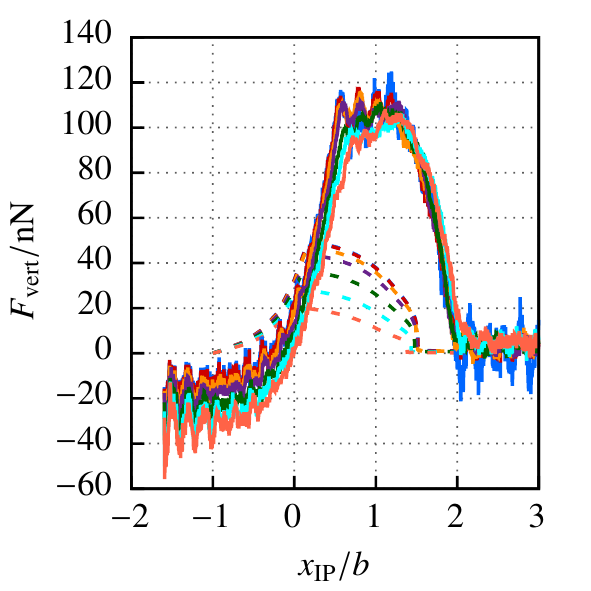} \\
	\begin{tabular}{c|c|c|c}
		\legline{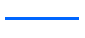} $0\%$ &\legline{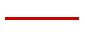} $0.1\%$  &\legline{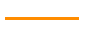} $0.5\%$ &\legline{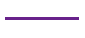} $1\%$ \\
		\legline{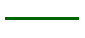} $2\%$ &\legline{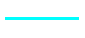} $3\%$ & \multicolumn{2}{l}{\legline{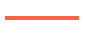} $4\%$ \quad \dots \quad defects}
	\end{tabular}\\
	solid: MD $\bullet$ dashed: PF
	\caption{Force comparison between MD and PF. The fracture toughness is estimated by minimizing the energy at the start and end of the MD simulations and making use of Eq.~\eqref{eq:1stAltEst}. The continuum forces underestimate the MD forces. The initial negative offset of $F_\text{vert}$ (solid lines) is due to the initial random perturbation of the atoms which results in reaction forces.}
	\label{fig:force_new}
\end{figure}

\section{Results and discussion}
\label{sec:paramident}

\subsection{Elastic material parameters}\label{sec:elMatParam}
For the continuum simulations, the \textsc{Young}'s modulus $E$ and the \textsc{Poisson} ratio $\nu$ have been determined as described in Sec.\ \ref{sec:methods_link} using \textsc{Tersoff-1994} and \textsc{LCBOP} in the MD homogenization. In Fig.~\ref{fig:elastHomo}, both potentials show a linear decrease of the \textsc{Young}'s moduli for small increasing defect percentages which is in line with predictions from elasticity theory for small holes in infinite plates~\cite{sneddon46}. Additionally, the variation increases for larger percentages which can be seen from the error bars indicating the $68.3\%$ quantile. For a perfect periodic single layer sheet, there is only one modulus for each direction without any variation. A variation does not occur before different defect configurations alter the moduli. The higher the defect percentage, the more configurations can be realized which yields an increasing variation of the moduli. At the same time, the \textsc{Poisson} ratios are almost constant up to approximately $5\%$ defects for the \textsc{LCBOP} potential. Up to $3\%$ defects, their variance is so small that it is not visible in the figure. The \textsc{Tersoff-1994} potential shows the same qualitative behavior but for smaller defect percentages. \par
For the crack propagation investigations, the \textsc{LCBOP} potential is chosen because it achieves more realistic values for the elastic parameters of graphene sheets. Of course, the homogenization procedure is not restricted to a specific choice of the potential. Furthermore, defect densities only up to 4\% are considered, since higher percentages yield partially unphysical, negative values for the elastic parameters. Possible reasons are the finite computational domain, which is not representative for large defect percentages anymore, or the fact that an increasing defect percentage will eventually lead to atoms which are not connected to any other surrounding atom. Additionally, isotropy is assumed for the considered defect range, which is reasonable considering the homogenization results. The average elastic parameters for $x_1$- and $x_2$-direction are noted in Tab.~\ref{tab:elastParam}. They are used within the continuum simulations. Besides the difference in the material parameters, the \textsc{LCBOP} potential incorporates long-range interactions while the \textsc{Tersoff-1994} potential does not. This is important to remember when it comes to the evaluation of the crack propagation simulations below.

\subsection{Fracture material parameters}
Fig.~\ref{fig:ensembleAverageDemo} shows a typical MD simulation result, where the surfing boundary condition was applied. The raw data from $250$ different simulations with 0\% defects, differing in the initial small random perturbation of the atoms, are plotted in grey. The blue lines depict the ensemble average. The vertical reaction force $F_\text{vert}$ is calculated as the average of the reaction forces for the upper and lower edge, where the boundary condition is applied.
A small negative offset at the beginning is observed, which is due to the small initial perturbation of the atoms and causes a reaction force before the actual loading begins. Before the initial crack propagates, the energy increases quadratically, which is consistent with elasticity. As soon as the crack starts to propagate, the force remains more or less constant, while the total energy increases almost linearly in $x_\text{IP}$. As soon as the crack has fully propagated through the specimen, the reaction force vanishes and the total energy remains constant. The final energy contains contributions due to the kinetic energy of the moving atoms, deformations and the surface along the crack. The resulting crack path is shown on the very left in Fig.~\ref{fig:ensembleAverageDemo}. As mentioned above, the crack path is not straight but wavy. \par
\subsubsection{Step-by-step energy minimization}
\label{sec:1stAlternative}
For each defect percentage, the fracture toughness has been estimated according to Eq.~\eqref{eq:1stAltEst}. Table \ref{tab:elastParam} lists the fracture toughness estimation depending on the defect percentage in the second to last row. The fracture toughness decreases with increasing defect density since the material is effectively weakened by the vacancies, which is physically evident.\par
Fig.~\ref{fig:force_new} compares the vertical reaction forces for the MD and PF simulations for different defect percentages. The PF simulations underestimate the MD forces by a factor of two to approximately five depending on the defects, which is very likely due to an underestimation of the dissipated energy: For the MD simulations, more than just one dissipation mechanism (bond breaking/crack surface increase) is contributing to a total dissipated energy. The large discrepancy between the MD and continuum results are the motivation to reconsider the method for the determination of the fracture toughness. \par
As already mentioned above, the PF simulation follows a quasi-static modeling approach, while the MD approach captures dynamic effects like the kinetic energy. On a continuum level, the kinetic energy from the atoms can be related to the temperature (=oscillation) and translational/rotational kinetic energy. Since the velocity of the applied surfing boundary condition yields quasi-static loading for the MD simulations, the latter contribution is negligible for the continuum simulations. However, it is observed, that the atoms' oscillation increases during cracking, which is equivalent to an irreversible increase of the temperature. Consequently, this phenomenon has to be accounted for in the continuum simulations but has been suppressed by the minimization procedure of this method. The second approach fixes this issue. 

\begin{figure*}[t]
	\centering
	\subfloat[Absolute reaction force]{
		\includegraphics{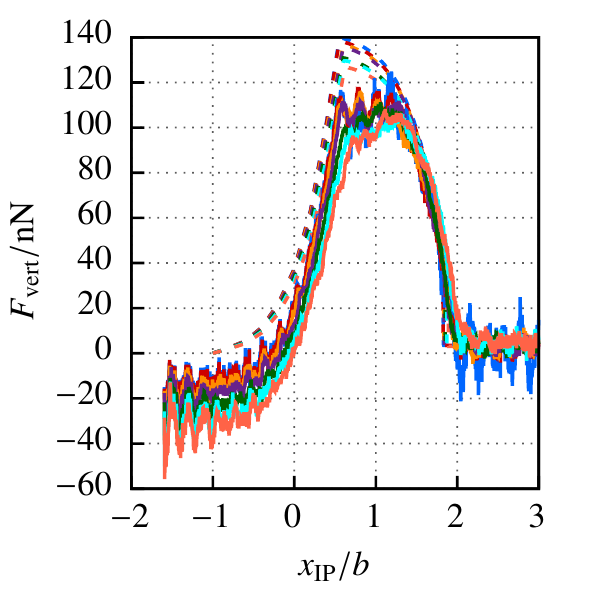}
	}
	\subfloat[Total energy]{
		\includegraphics{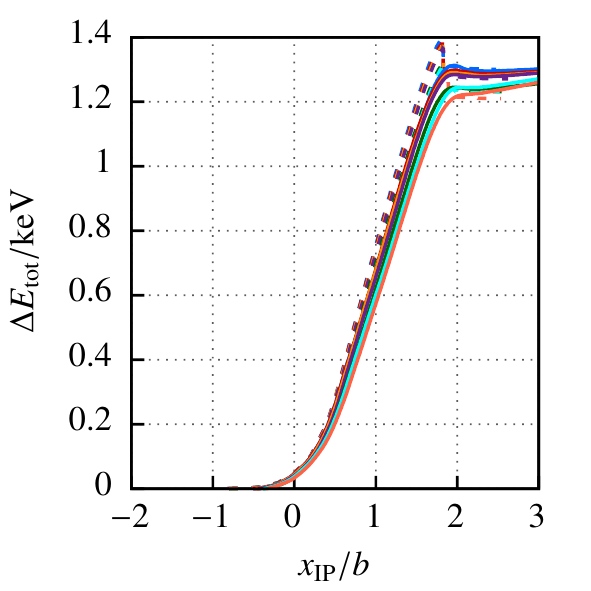}
	}\\
	\begin{tabular}{c|c|c|c|c|c|c}
		\legline{2} $0\%$ &\legline{3} $0.1\%$  &\legline{4} $0.5\%$ &\legline{5} $1\%$ &\legline{6} $2\%$ &\legline{7} $3\%$  &\legline{8} $4\%$\quad \dots \quad defects
	\end{tabular}\\
	solid: MD $\bullet$ dashed: PF
	\caption{Absolute reaction force (a) and total energy (b) of defective single layer sheets. The lines resemble the ensemble average over $250$ samples at each point in time. For the determination of the fracture toughness, the energy at $x_\text{IP}=2$ is considered to be the energy after complete rupture. The initial negative offset of $F_\text{vert}$ (solid lines) is due to the initial random perturbation of the atoms which results in reaction forces.}
	\label{fig:MDForceEnergy}
\end{figure*}

\subsubsection{Total energy difference}
\label{sec:2ndAlternative}
The previous approach to calculate the fracture toughness neglected certain contributions to the dissipated energy. In order to resolve these issues, the difference in the total energy is used to determine the fracture toughness: Even if the continuum approach does not explicitly account for a temperature, its increase and influence can be included in the fracture toughness by means of an effective fracture toughness value. In other words, an increase of the crack surface goes in line with an increase of other irreversible energy contributions, which are now all accounted for by the crack energy. The right diagram in Fig.~\ref{fig:MDForceEnergy} depicts the increase in total energy for the MD simulations (solid lines).
Table \ref{tab:elastParam} lists the values for the second fracture toughness estimation in the last row.
As in the previous section, the fracture toughness decreases with increasing defect density.\par
The right diagram in Fig.~\ref{fig:MDForceEnergy} compares the total energy for the MD and PF simulations. At around $x_\text{IP}/b\approx1.8$, the specimen abruptly fails. This point in time is clearly visible in the energy plots where the dashed lines exhibit a kink and strong decrease to their final level. The sudden decrease is due to the elastic relaxation of both pieces of the broken specimen and the inertia effects which are neglected in the quasi-static approach. The final energy level of the PF simulations perfectly compares to the MD results, which is expected since the energy from MD simulations was used to determine the fracture toughness for the PF simulations. Again, the overestimation of the fracture toughness, Eq.~\eqref{eq:gcnum}, is compensated. \par
The left diagram in Fig.~\ref{fig:MDForceEnergy} compares the vertical reaction forces for the MD and PF simulations for different defect percentages. Very good agreement between the MD and PF simulations is achieved compared to the previous approach which incorporated a step-by-step energy minimization. As the increase in total energy served as input parameter for the determination of the fracture toughness, all dissipation phenomena of the MD simulations are captured in an effective manner within the quasi-static PF simulations.

\subsection{Discussion on the regularization length scale}
\label{sec:lengthScaleDisc}
\begin{figure*}[t]
	\centering
	\subfloat[Absolute reaction force]{
		\includegraphics{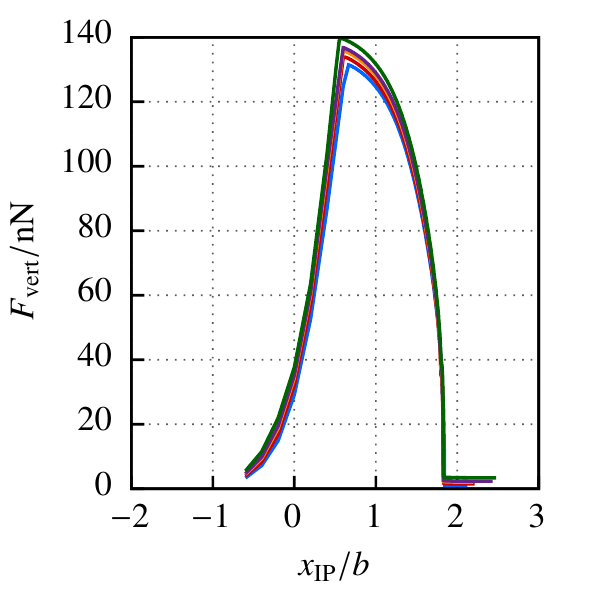}
	}
	\subfloat[Total energy]{
		\includegraphics{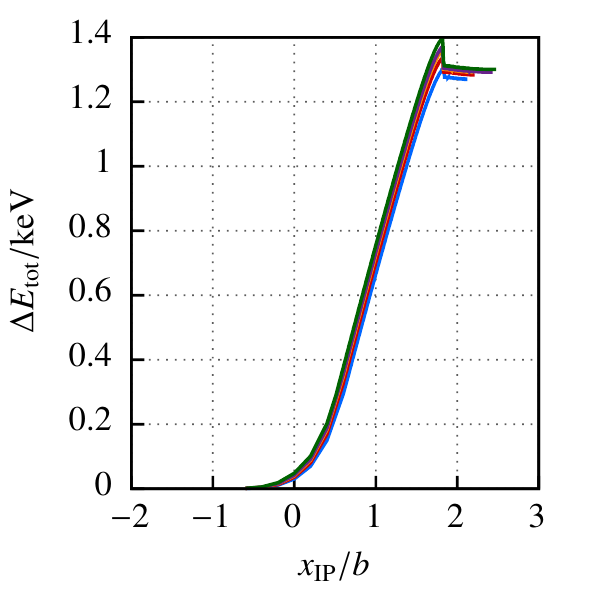}
	}\\
	\begin{tabular}{l|l|l|l|l}
		\legline{2} $l_\text{c}=0.045b$ &\legline{3} $l_\text{c}=0.03b$  &\legline{4} $l_\text{c}=0.0225b$ & \legline{5} $l_\text{c}=0.02b$ &\legline{6} $l_\text{c}=0.015b$
	\end{tabular}
	\caption{Absolute reaction force (a) and total energy (b) of the continuum approach: For the PF simulations, the length scale $l_\text{c}$ is varied to study its influence on the results. The only criterion for the initially chosen value $l_\text{c}=0.015b$ was, that it should be in the same order of the bond length. This already yields a good agreement between the MD and PF approach. As expected, smaller length scales yield a higher maximum force and vice versa. The small force offset at the very end of the simulations is due to the residual stiffness. The energy at the end of the simulations ($x_\text{IP}\geq2$), depicted in the right diagram, is almost independent from the length scale. Very small deviations occur, because the discretization is the same for each case and larger length scales are better approximated than smaller ones.}
	\label{fig:lcVar}
\end{figure*}
For the sake of completeness, the influence of the PF length scale on the results is studied. Patil~\cite{patil_comparative_2016} and Padilla~\cite{padilla_3d_2017} propose a density-based determination of the characteristic length scale. They calculate the normalised density along a line perpendicular to the crack and match the resulting curve to the analytical solution of the PF.\par
In this work, the length scale is assigned an arbitrary value which is on the order of the distance between the atoms. As shown above, this arbitrariness yields results which are in good agreement with the MD simulations. Now, the length scale is varied and the effect on the results is studied. \par
The left diagram in Fig.~\ref{fig:lcVar} compares the force curves for different length scales to the original value $l_\text{c}=0.015b$ (green curve). As expected from a PF model, the maximum force decreases for an increase of the length scale. It is noted, that only values larger than the original values are investigated, because of the computationally very expensive discretization, which would have been necessary, if smaller values had been considered.\par
The right diagram in Fig.~\ref{fig:lcVar} compares the energy for a variation of the length scale. The curves are matching perfectly well, which is expected from the PF formulation: The length scale has no effect on the energy dissipation. The very small deviations are due to the difference in length scales, where larger values are better approximated by the finite element mesh, which has the same characteristic length $h$ for each simulation. It is noted, that the overestimation of the fracture toughness, Eq.~\eqref{eq:gcnum}, is compensated. Based on this information, the length scale variation could potentially be used to verify the value, which had been determined using a density-based approach~\cite{patil_comparative_2016} by comparing the force curves or to find another value for $l_\text{c}$ to yield an even better agreement between the MD and PF simulation results. This goes, however, beyond the scope of this work.


\section{Conclusions and limitations}
\label{sec:conclusion}
In this article, a combined modelling approach featuring MD and PF simulations has been used to study the influence of defects on crack propagation in single graphene sheets.
All material parameters were determined by means of MD simulations using a homogenization scheme for the elastic constants ($E$ and $\nu$) and crack propagation studies for the fracture toughness ($\mathcal{G}_\text{c}$). For the determination of $E$ and $\nu$, two different energy potentials (\textsc{Tersoff-1994} and \textsc{LCBOP}) were used and the results were compared. Two alternatives were presented for linking the fracture toughness between both scales based on energetic criteria. Both approaches go without a global parameter fitting but use energetic relations instead, which is a novelty. In both cases,
the fracture toughness decreased with increasing defect density. For \textsc{LCBOP} we obtain $\mathcal{G}_\text{c}\approx5.4\;\text{nN}$ in case of pristine graphene (0\% defects) applying the \textit{step-by-step energy minimization}. This value is in excellent agreement with recent measurements reported in Ref.~\cite{Zhang2014} where $\mathcal{G}_\text{c}=5.4\;\text{nN}$ -- for an assumed ``thickness'' of $\SI{0.34}{\nano\meter}$ -- was found. Compared to the second presented method -- \textit{total energy difference} -- this value is much smaller, cf. Tab.~\ref{tab:elastParam} first column. This discrepancy can be explained with the following argument: The higher effective fracture toughness does not only include the dissipation contributions due to fracture, but also all other dissipation. This, however, is not comparable to the experiments anymore. Indeed, the value obtained using the \textit{step-by-step energy minimization} is the one which only contains fracture contributions and thus, is the right one when it comes to a comparison. A remedy to solve the discrepancy for the second approach would be the explicit involvement of the temperature field within the continuum simulations in order to account for the atomistic oscillations on MD level in terms of a temperature change on continuum level. This is to be investigated in the future. 

Further, it was shown that quasi-static phase-field simulations are able to capture highly transient processes in an effective manner which are happening on short time scales in MD simulations. Finally, the results of the PF length scale study revealed only small dependence of the results on the regularization parameter. Altogether, a rigorous identification of three material parameters on the MD scale and application on the continuum scale was presented for defective graphene sheets. \par
Open questions concern a rigorous identification of the PF length-scale with a physical length. Here, the density-based approach \cite{patil_comparative_2016} is a good starting point. Furthermore, the presented PF model can be extended to large deformations to see whether the small strain approach is admissible. Additionally, it would be of great interest to explicitly account for the temperature in the continuum simulations and determine arising material parameters in the same manner. The foremost challenge is the extension of the PF model to represent a three-dimensional defective stack of few graphene sheets, which introduces anisotropy and heterogeneities along the third dimension \cite{teichtmeister2017,bleyer2018}.

\section*{Data Availability}
The raw/processed data required to reproduce these findings cannot be shared openly, yet on request, at this time as the data also forms part of an ongoing study. 
\section*{Acknowledgements}
The authors acknowledge the computational resources from the TU Dresden centre for information services and high performance computing (TUD ZIH).




\bibliographystyle{elsarticle-num}
\bibliography{literatur}

\end{document}